\documentstyle[11pt,paspconf]{article}

\begin{document}

\title{A hundred years of science at the Pic du Midi Observatory}
\author{Emmanuel Davoust}
\affil{Observatoire Midi-Pyr\'en\'ees, 14 Avenue Belin, F-31400
Toulouse, France}

\begin{abstract}
The Pic du Midi Observatory, situated at an altitude of 2890 meters, was 
always very hard to reach, and living there was difficult. Its history is a 
lesson in courage. It is also a lesson in creativity, because astronomers took 
advantage of the remarkable quality of the site in many ways, to study planets 
and, later, to prepare for the Apollo missions.  They also invited 
geophysicists, botanists, and cosmic ray physicists to conduct experiments 
there, and the Observatory became a successful center for multidisciplinary 
studies. The heroic period, which ended in 1952, is reviewed through the 
accomplishments of four directors, C\'elestin Vaussenat, Emile Marchand, 
Camille Dauz\`ere and Jules Baillaud. 
\end{abstract} 


\section{Introduction}

The Pic du Midi Observatory is located at an altitude of 2890 meters (9470 
feet) on an isolated summit of the Pyr\'en\'ees range which separates France 
from Spain.  Its largest instrument is the 2-meter Bernard Lyot telescope, 
which is mainly used for CCD imaging and spectroscopy; it is also equipped 
with a Nicmos infrared camera.  Next is a 1-meter telescope, also equipped 
with CCD cameras, mainly used for planetary research.  Several other domes 
house coronographs and other instruments for solar research; the site is one 
of the best in the world for solar observations, with long periods of stable 
and excellent (down to 0.3 arcsec) seeing in the morning and early afternoon.  
This Observatory is also open to amateur astronomers, and a 60-cm reflector is 
exclusively reserved for their projects.  The summit also houses an important 
communication, radio and TV transmission center with a tall antenna. 

The Pic du Midi Observatory has acquired its world wide fame from its 
remarkable astronomical achievements : the mapping of planetary surfaces, the 
determination of the period of rotation of Venus, the preparation of the 
Apollo Moon landings.  Astronomy is presently the most important field 
of research conducted at this Observatory, but it was initially a 
meteorological station and, over the years, it has hosted research in many 
other scientific domains, such as Earth magnetism, atmospheric physics, 
seismology, natural radioactivity, glaciology, cosmic ray physics, botany, 
and, to a lesser extent, artificial radioactivity, physiology and medical 
research. 

A review of the scientific results obtained at the Pic du Midi Observatory 
would certainly be very interesting, but, in this historical account, I will 
emphasize the scientists rather than the science. Living permanently and 
conducting research at the Pic du Midi Observatory was undoubtedly difficult 
in view of the isolation and the altitude of the site. In the 1930's, one of 
the staff members used to say that he lived a six hour walk away from France.  
Until the cable car was built in 1952, it took a lot of dedication to live 
permanently up there, at the expense of family life, as one could not go home 
every day or even every week-end, and of health.  Poor heating and inadequate 
ventilation of the living quarters caused chronic bronchitis, and the poor 
diet, mainly composed of preserves, salted pork and the like, often caused 
digestive problems. 

The history of the Pic du Midi Observatory actually started about 1000 feet 
below the summit, at the Sencours pass, where a temporary meteorological 
station was established in 1873.  The two leading figures of the project were 
Charles Champion du Bois de Nansouty, a retired army general, and 
C\'elestin-Xavier Vaussenat, a construction engineer by trade. While Nansouty 
spent about eight years at the Sencours pass conducting daily routine 
meteorological observations, Vaussenat criss-crossed the country giving public 
conferences to raise money and find sponsors for their project, which was an 
entirely private initiative. 

Construction of the Observatory at the summit started in 1878 and lasted 4 
years, because it could only be conducted during the 4 months or so (mid-July 
to early or mid-October) that the summit was free of snow and easily 
accessible. The main concern of the founders was to protect the workers from 
lightning which frequently hit this isolated peak.  Several lightning rods 
were installed, and linked to the nearby lake of Oncet by a metallic cable 
about a mile long. 

The Observatory was inaugurated in August of 1882. At that point, the founders 
of the Observatory realized that they would never have enough funds to 
maintain their establishment.  They thus offered to donate the Observatory to 
the State, on the condition that it would pay their debts and provide a yearly 
allowance of 30 000 francs, a large sum at the time, for the salary of the 
director and the maintenance of the site and its activities.  Under the 
pressure of prominent politicians and scientists lobbied by Nansouty, the 
State accepted the donation; the establishment became a National 
Observatory supervised by the Central Bureau of Meteorology in Paris and 
Vaussenat became its first director.  Nansouty retired to the city of Dax and 
never went back to the Pic du Midi. 

\section{C\'elestin Vaussenat (1882 - 1891)}

Being an engineer and not a scientist, Vaussenat devoted most of his time and 
effort to expanding the Observatory.  He razed most of the summit to make 
place for several terraces, built a storage house (later called the Vaussenat 
building) and the so-called ``blockhaus" on top of which all the meteorological 
instruments were housed.  Since the terraces were covered with several meters 
of snow during most of the winter season, a tunnel had to be constructed in 
the rock to facilitate the access to the blockhaus. 

Vaussenat invited scientists to conduct experiments at the Observatory.  In 
November 1882, two astronomers from Paris came to observe the transit of 
Venus over the Sun.  Because of bad weather, they did not manage to bring 
their heavy equipment to the summit and had to settle at Sencours pass.  On 
December 6, the day of the transit,
the weather was clear at the summit, but overcast at Sencours 
pass.  At that time occurred the only major tragedy in the history of the 
Observatory, when the porters were caught in an avalanche which caused the 
death of three of them. After that, the porters were extremely cautious and 
would not climb to the summit when the risk of avalanche was high, usually 
leaving from the valley well before sunrise.

Other visitors of that period included physicists who measured the carbon 
monoxide and ozone content of the air in 1881 and the next two summers, two 
astronomers from Paris in 1883, who tested the site and found it exceptionally 
good, army geodetists in 1884 who were making a new map of France, Jules 
Janssen from Meudon Observatory who observed the Sun in 1887,  a medical 
doctor who studied the effects of altitude on various animals (dogs, chickens, 
cats, rabbits, etc.) in 1890, and two astronomers from Lyons Observatory, also 
in 1890. 

At the request of the Central Bureau of Meteorology, Vaussenat conducted an 
expensive and elaborate project between 1883 and 1885.  The purpose of this 
``Lemstr\"om experiment" was to produce artificial aurorae.  Selim Lemstr\"om 
was a Finnish physicist who successfully carried out such an experiment in the 
North of Finland, and reported his results to the French scientific community 
(Lemstr\"om 1884). Vaussenat purchased 200 long wooden sticks treated against 
weathering, planted them evenly over 530 m$^2$ on the summit and joined them 
with a mesh of metal wire studded with 10800 metal points.  The instrumental 
setup did not produce the expected result; on the other hand, it efficiently 
attracted bolts of lightning which forced Vaussenat to buy himself a new suit 
and watch. 

\section{Emile Marchand (1882 - 1914)}

Vaussenat had a heart problem at the summit in November of 1891, was carried
down to the valley by two porters and died in Bagn\`eres a week later.  After a 
fierce tug of war between the Central Bureau of Meteorology, Paris Observatory 
and other institutes, Emile Marchand, a seasoned astronomer and geophysicist 
from Lyons Observatory, who had visited the Pic du Midi Observatory two 
summers before, became the second director of the Pic du Midi Observatory.  

He immediately took on the task of improving and expanding the scientific 
observations and mesurements conducted at the Observatory. During the 22 years 
of his directorship, he collected and partly analyzed a remarkable set of 
routine daily observations made by his staff at the summit, of astronomy, 
meteorology (including a map of the cloud cover), static electricity, the 
horizontal and vertical components of the Earth's magnetic field, seismology.  
During that period, he published 35 papers on meteorology, 22 on geophysics, 
20 on astronomy, 6 on solar-terrestrial relations, and 9 on botany.  
Unfortunately, he published most of his papers in the journal of a local 
scientific society and in obscure conference proceedings, so his work is still 
waiting for its readers. 

The astronomical observations, a daily map of sunspots, a study of the 
aspect of Venus (to determine its period of rotation), the search for an 
atmosphere on the Moon, occultation events of the satellites of Jupiter, were 
all made by his assistant Sylvain Latreille, using an 8 inch refractor of 
mediocre quality donated by Paris Observatory.  Marchand acquired a 
spectroheliograph, but the instrument was defective and he never managed to 
make it function properly.  He made detailed plans for installing a polar 
siderostat, an instrument well suited for a permanently snowed in site, but 
his expensive project never materialized, for lack of support from Paris.

Marchand resided in the town of Bagn\`eres, 20 km away, where 
he conducted the same kinds of observations as at the top of the mountain.
Since he would also conduct these measurements on his way up to the summit, 
this would give him detailed information on the variations with altitude
of all the physical parameters.  He was in daily contact with the Observatory 
through a private phone line, but the connection was so bad that they often 
communicated by Morse signals.  The line was also cut off altogether several 
times a year by snow, falling branches, avalanches, or grazing sheep, and 
Marchand had to send someone to find and repair the damage.  Since the line 
was made of copper, part of it was stolen on two occasions. 

Marchand spent a lot of time on administrative duties and on practical 
tasks linked to the maintenance and supply of the Observatory in food, fuel 
and other indispensable items.  During the brief period of summer, when there 
was no snow, a permanent train of mules (carrying a load of up to 100 kg each) 
would carry to the summit 10 to 15 metric tons of coal, 2 tons of potatoes, 50 
12-gallon barrels of wine, as well as wood, tiles, drinking water, 
preserves, etc. The rest of the year, the Observatory would be supplied with 
fresh meat, bread, cheese and vegetables by 2 or 3 porters, who, weather 
permitting, made the 5 to 8 hour climb on Sunday morning, the day after market 
day in Bagn\`eres.  These expensive means of transport would double the cost 
of all the material used at the summit.  Furthermore, since the transport 
would often be delayed or even cancelled when snow fell (because of the 
increased risk of avalanches), the food would go bad and be lost. 

At the turn of the century, the Central Bureau of Meteorology asked Marchand 
to build a botanical garden at the summit, in order to study the behavior of 
plants at high altitude.  Joseph Bouget, a modest gardener from Bagn\`eres,
took charge of the project, and over the years he and others published over 30 
papers on the botanical experiments conducted at that garden; he became an 
expert who was consulted by the most renowned botanists of France.  In the 
mid- and late-1930's he managed to produce potatoes from seeds (Bouget 1935), 
which is apparently quite a feat. But the ministry consistently refused to give 
him a permanent position at the Observatory, and he was always paid by 
expedients. 

Also at the beginning of the century, Benjamin Baillaud, then director of 
Toulouse Observatory, decided to build an astronomical telescope at the Pic du 
Midi Observatory.  Baillaud was often on the inspection team mandated by the 
Central Bureau of Meteorology to inspect the Observatory every summer, so he 
was familiar with the site and well aware of its advantages.  But first he 
wanted to establish for himself the high quality of the site. He had a 
temporary dome installed on the summit, and he and members of his staff 
spent several summers doing test astronomical observations with several small 
telescopes. They concluded that the images were indeed often remarkable
(Baillaud \& Bourget 1903). Baillaud obtained funds for the construction of a 
50-cm reflector and a house for the visiting astronomers, and construction 
began in 1904.  In the summer of 1906, the dome was ready and the telescope 
built in Paris by Gautier was shipped to the foot of the mountain.  The 22 
crates, weighing between 600 and 1600 pounds, were then transported by oxen to 
the Tourmalet pass, and from there to the summit with great difficulty by a 
dozen soldiers belonging to a nearby artillery regiment.  In fact, they only 
managed to reach the Sencours pass before the first snowfall, and the 
telescope parts only reached the Observatory the next summer. 

It took another summer to install the telescope properly, and it was only in 
1909 that it became fully operational.  It was first used by Count Aymar de la 
Baume Pluvinel, a private astronomer from Paris, and his assistant Fernand 
Baldet, to observe Mars in the fall of 1909.  The excellent images that they 
obtained allowed them to claim that the canals of Mars were not visible
(de la Baume Pluvinel \& Baldet 1909).  The next year was that of Halley's 
comet. Unfortunately, it was visible in May, the worst time of the year for 
observing at the Pic du Midi Observatory. Henri Godard from Bordeaux 
Observatory and Gaston Millochau from Meudon Observatory spent over three 
weeks without opening the dome.  One of the rare forms of entertainment was a 
record player with half a dozen records, and, by the end of their stay, they 
knew pretty well all of them by heart.  The same year, de la Baume and Baldet 
returned to the Pic du Midi for observing Saturn, and ran into the same bad 
weather as their colleagues in the spring, so the Observatory immediately 
acquired the ill-founded reputation of being permanently clouded over, and was 
very seldom used by French astronomers.  The only one who persistently 
observed at the 50-cm telescope was Jules Baillaud, the son of Benjamin 
Baillaud, and his thesis is based on these observations (Baillaud 1924). 

The new telescope, managed by Toulouse Observatory, was a thorn in 
Marchand's side from the very start. He felt it as an intrusion on his turf; 
so did his staff at the summit, and friction between them and the workers and 
visitors from Toulouse continuously erupted. 

Emile Marchand died in March of 1914, and, before a new director was 
appointed, war broke out. The anxiety felt in these difficult moments is 
discernable in a letter of the caretaker director :``I have to discontinue my 
work at the Observatory because I have been mobilized.  The office will be 
closed as of today. (...) The two assistants, also mobilized, left the 
Observatory yesterday. Mr. Labayle, observer, must leave in two weeks.  At the 
summit there will be Mr. Latreille, {\it all alone}."  And alone at the summit 
he remained for 14 months without interruption, until someone was found to 
replace him temporarily, because he would not allow any gap in the daily 
meteorological observations.

This was the occasion to put all the instruments and staff at the summit under 
a single authority (that of Toulouse Observatory) and Pic du Midi Observatory 
merged with Toulouse Observatory in 1915.  Joseph Rey, a Navy officer, was 
nominated vice-director of Pic du Midi Observatory in 1915, but only took 
office in late 1917.                        

\section{Camille Dauz\`ere (1920 - 1937)}

For several reasons, Rey soon resigned from his position and was replaced by 
Camille Dauz\`ere, a physicist, in 1920.  The latter immediately started 
renovating the buildings which had considerably suffered from lack of 
maintenance during the war. But it was already a bit late, as part of the 
northern terrace collapsed in July, 1922. This was paradoxically a lucky 
event, as it helped Dauz\`ere emphasize the impending catastrophe if nothing 
was done about the poor state of affairs, and politicians on all sides 
mobilized to save the Observatory.  This mobilization actually happened again 
in the early 1990's when the Ministry decided to close the Observatory.  As a 
result, Dauz\`ere received adequate funding for over ten years to renovate the 
buildings and build new ones.  

French astronomers did not show the same eagerness to renovate astronomy at 
the summit. Henri Deslandres, director of Meudon Observatory, was mandated in 
1922 to inspect all French Observatories.  After visiting the Pic du Midi 
Observatory he recommended closing it.  So Dauz\`ere resolutely set his course 
toward geophysics to save the establishment, and for the next 15 years this was 
the main field of research conducted at the Pic du Midi, together with botany. 
Joseph Devaux did his thesis on the thermal behavior of snowfields 
and glaciers (Devaux 1933), and participated in several expeditions to 
Greenland with Charcot. Hubert Garrigue did his thesis on natural 
radioactivity in the mountains (Garrigue 1936). Dauz\`ere himself became 
interested in thunderstorms and hail, using the remarkable database on the 
location of lightning impacts in the region, collected by Joseph Bouget over 
twenty years, and expanding it by sending out forms to be filled out by 
witnesses of lightning and hail storms. He thus became a nationally famous 
expert on the subject. 

A gradual shift toward astronomy occurred in the 1930's, when Bernard Lyot 
invented the coronograph and experimented it at the Pic du Midi between 1930 
and 34 (Lyot 1939).  This revolutionary instrument allowed the solar corona to 
be observed outside eclipses.  But it took the talent of Lyot to produce a 
diffusion-free instrument and the quality of the atmospheric conditions at Pic 
du Midi to make the experiment successful.  In 1935, Jules Baillaud, a strong 
advocate of astronomy at the Pic du Midi, became president of the society 
``Les amis de l'Observatoire du Pic du Midi".  In this quality, he proposed a 
project for improving the existing 50-cm telescope built by his father and for 
building a new one.  Finally, in 1936, the ministry appointed an astronomer 
rather than a geophysicist to replace Joseph Devaux who disappeared with 
Charcot off the coast of Iceland.  This was Henri Camichel, resolutely decided 
to observe even in winter, and one of the members of the team that conducted 
very successful planetary observations at the Pic du Midi in the 1940's
and 1950's.

\section{Jules Baillaud (1937 - 1947)}

When Camille Dauz\`ere retired in 1937, nobody volunteered to take his 
position.  The ministry then decided that, if no candidate for the position of 
vice-director was found, the Observatory would be shut down.  At that point 
one man stepped forward, because he could not bear the idea that this 
Observatory whose development his father had so much contributed 
to would be closed. This man was Jules Baillaud.  At the time, he was 63 years 
old, thus soon to retire, and head of La Carte du Ciel, the sky survey of the 
time, at Paris Observatory.  Furthermore, as everybody knows, Paris is the 
center of the world, so why go and live in some small town in south-western 
France? He nevertheless reluctantly accepted the position, on the condition 
that he would share his time between Paris and Bagn\`eres, and that a 
responsible person in Bagn\`eres would be in charge and do all the paper work 
in his absence. That person was Charles Taule, a former meteorologist at the 
summit and primary school teacher, who would contribute a lot to the survival 
of the Observatory during the war years. 

Jules Baillaud soon embarked upon an ambitious project of renovating astronomy 
at the Observatory, and of improving living conditions by installing a cable 
car and replacing the batteries, an unreliable and weak source of electric 
current, by an electric line from the valley.  But he had hardly time to do 
much before war broke out and all the staff of the Observatory was again 
mobilized and sent to the front.  

One would think that an old man, who didn't really want to be there, alone in 
this empty Observatory, in a period when the whole country was in great 
turmoil, would just give up.  Well, he did just the contrary.  This is perhaps 
a typical case of Gallic pride; in the face of defeat and humiliation, France 
has to stand up and show that, in Baillaud's own words, ``its genius and 
influence are not extinct". This is very similar to what happened after 1870 
and the defeat of the French by the Prussian army : in the ten following 
years, French astronomers built six provincial observatories with funds from 
the State, as well as two private ones. 

Baillaud first sought to renovate the existing 50-cm refractor, whose mirror 
was of poor quality.  He negociated with Ren\'e Jarry-Desloges, a wealthy 
amateur astronomer and owner of a good 50-cm objective, but the latter was not 
very eager to lend his precious lenses; furthermore, he did not believe one 
could do good photographic work with it.  The imminence of the 1941 opposition 
of Mars forced the astronomers to find a solution rapidly. They borrowed the 
38-cm objective of Toulouse Observatory, and they made excellent observations 
of Mars which convinced them that a permanent solution had to be found. 

Lyot then thought of the 60-cm objective of the coud\'e telescope of Paris 
Observatory.  But its focal length was 18 meters, three times that of the 50-cm 
telescope, which required that the light rays be folded twice, with two flat 
mirrors. The telescope would then in effect become a refracto-reflector, a
system already successfully tested by the Swiss astronomer Emil Schaer in the 
beginning of the century.  This was quite feasible, since the existing 
Pic du Midi telescope was composed of two tubes, one for the 50-cm reflector 
and one for the 23-cm refractor used for guiding.  The partition between the 
two tubes was knocked out and three bearings for the objective and two flat 
50- and 30-cm diameter mirrors ordered from an instrument factory near Vichy.
The pieces were manufactured and shipped amidst incredible difficulties of 
production, communication and transport. This was in the fall of 1942, when 
the Allied forces landed in North Africa and the Germans occupied the Southern 
part of France to prevent any landing there. 

The refracto-reflector, later called ``lunette Baillaud" was very successfully 
used until the early 1960's, to measure very accurately the diameter of 
planets, to map the surface of Mercury, to study the spots on the surface of 
Jupiter's satellites (Dollfus 1961).  One had to wait for observations from 
space to get better results. 

Baillaud then turned to his other project, the construction of two new 
telescopes, a 85-cm Schmidt telescope for wide-field photography, and a
150-cm long-focus reflector for high-resolution studies.  The design of the 
latter, mainly due to Lyot, was very bold and revolutionary, well ahead of its 
time. It was to be a domeless telescope, with a closed tube filled with helium 
and refrigerated from the outside.  The absence of a dome would eliminate any 
local turbulence due to such a structure. Completely retractable domes are 
becoming a standard for new generation telescopes such as the VLT.  But the 
telescope had to be very rigid, in order to operate with high winds, of up 
to 20 m/s; the frost and snow presented another serious difficulty for the 
design. The helium in the tube and the external refrigeration would improve 
the temperature stability and thus the seeing.  Several new techniques for 
maintaining the shape of the mirror during pointing were considered. 

The mirror would not be made of glass, but of steel. There was a very simple 
practical reason for that, namely that the Saint-Gobain factory's ovens had 
been destroyed by the Germans and could not produce blanks of the required 
size.  The advantage of steel is its rigidity and thermal stability. 
The disadvantage of steel is its low power of reflectivity, less than 65\%. 
Baillaud made thorough investigations for improving the reflectivity, by a 
layer of chrome or of various alloys.  Another problem was that nobody knew 
how to polish steel to the required shape.  

The feasibility studies started in January of 1943, and were done by Messier, 
an aircraft landing gear manufacturer, out of work because the Germans forbade 
any sensitive R\&D in occupied France.  The advantage of this company over 
traditional telescope makers was its lack of preconceived ideas on how to 
build a telescope, and thus its openness to new ideas.  It met considerable 
material difficulties, especially in June of 1944, when the country was 
completely disrupted by the Allied landing in Normandy. No material of any 
kind was available for building models, no wood, screws, nails, glue or 
soldering material. Mail would take weeks to arrive, and many higher ranking 
civil servants were put in prison by the new government, in effect paralyzing
the country.

It is paradoxically the end of the war that put an end to the project.
Messier returned to its more profitable work on aircraft R\&D, Baillaud soon 
retired, Lyot died prematurely in 1952, but the main reason why this project 
never materialized was the stubborn opposition of Andr\'e Danjon, director of 
Paris Observatory and {\it de facto} boss of French astronomy. He had decided 
once and for all that no good images could be obtained at high altitude, and 
he wanted to develop the competing site of Haute-Provence and provide it with 
a large telescope.  The large telescope project at Pic du Midi only went ahead 
in the sixties, after Danjon retired, and, by that time, the design was 
completely reconsidered by the entirely new team of astronomers. 

\section{The present and the future}

The end of the war nevertheless saw the completion of Baillaud's other 
projects for improving conditions at Pic du Midi Observatory.  A high voltage 
power line soon reached the summit, allowing expensive cosmic ray experiments 
to be conducted, and a cable car was inaugurated in 1952, in effect creating
a tight border between ``before" and ``after", between those who had known the 
heroic period, when it took many hours of difficult climbing to reach the 
summit, and those who arrived up there in a suit and tie and with a briefcase,
as if they had just stepped out of the subway.

With these essential improvements, research at the summit expanded 
spectacularly, with up to 10 or 20 people at the Observatory at a time, 
several research teams simultaneously at the various telescopes and 
instruments, doing exceptional planetary observations and solar research, 
monitoring the solar corona every day, visitors from Manchester (England) for 
the study of cosmic rays and later for the preparation of the Apollo Moon 
landings. 

As for the future, it is not as bright as one might expect. In the present 
context of economic crisis, the ministry again threatened to close the 
Observatory, and the only way out was to lend two thirds of its surface to 
private enterprise, which in turn will open it to the public, with a science 
museum, astronomy shows, a restaurant, shops, etc.  The site will be 
accessible all year round and 150000 visitors are expected annually, starting 
in 1998.  My hope is that, while this showcase of astronomy will be a drawback 
for astronomers at the Pic du Midi, it will entice the taxpayers to contribute 
more funds for French astronomy in the future.

\acknowledgments

The 14 photographs illustrating this paper are available in gif format at 
http://www.obs-mip.fr/omp/umr5572/patrimoine/asp.html.  
This is an invited lecture at the History session (Astronomy from difficult 
places) of the 109th Meeting of the Astronomical Society of the Pacific 
jointly with the Astrophysics From Antarctica Symposium.
It will appear in Astrophysics From Antarctica - Scientific Symposium 
Conference Proceedings (ASP Conference Series).

\end{document}